\documentclass[]{aa}
\usepackage{times}

\usepackage{graphics,psfig,epsf,rotate}

\def\FUSE{{\it FUSE}}

\def\arcsec{\ifmmode '' \else $''$\fi}
\def\arcmin{\ifmmode ' \else $'$\fi}
\def\arcsecpoint{\ifmmode ''\!. \else $''\!.$\fi}
\def\arcminpoint{\ifmmode '\!. \else $'\!.$\fi}
\def\cc{\ifmmode {\rm cm}^{-3} \else cm$^{-3}$\fi}
\def\cl{\ifmmode {\rm cm}^{-2} \else cm$^{-2}$\fi}
\def\micron{\ifmmode \mu{\rm m} \else $\mu$m\fi}
\def\kms{\ifmmode {\rm km\,s}^{-1} \else km\,s$^{-1}$\fi}
\def\Hubble{\ifmmode {\rm km\,s}^{-1}\,{\rm Mpc}^{-1}
        \else km\,s$^{-1}$\,Mpc$^{-1}$\fi}
\def\ergsec{\ifmmode {\rm ergs\;s}^{-1} \else ergs s$^{-1}$\fi}
\def\ergscm{\ifmmode {\rm ergs\,s}^{-1}\,{\rm cm}^{-2}
          \else ergs\,s$^{-1}$\,cm$^{-2}$\fi}
\def\ergscmA{\ifmmode {\rm ergs\,s}^{-1}\,{\rm cm}^{-2}\,{\rm \AA}^{-1}
          \else ergs\,s$^{-1}$\,cm$^{-2}$\,\AA$^{-1}$\fi}
\def\ergscmHz{\ifmmode {\rm ergs\,s}^{-1}\,{\rm cm}^{-2}\,{\rm Hz}^{-1}
          \else ergs\,s$^{-1}$\,cm$^{-2}$\,Hz$^{-1}$\fi}
\def\Msun{\ifmmode M_{\odot} \else $M_{\odot}$\fi}
\def\Lsun{\ifmmode L_{\odot} \else $L_{\odot}$\fi}
\def\qo{\ifmmode q_{0} \else $q_{0}$\fi}
\def\Ho{\ifmmode H_{0} \else $H_{0}$\fi}
\def\hi{H {\sc i}}

\def\lya{Ly$\alpha$}
\def\lyb{Ly$\beta$}
\def\lyg{Ly$\gamma$}

\def\ciii{C\,{\sc iii}}
\def\niii{N\,{\sc iii}}
\def\civ{C\,{\sc iv}}
\def\nv{N\,{\sc v}}

\newcommand{\ovi}{O~{\sc vi}}
\newcommand{\ovii}{O~{\sc vii}}
\newcommand{\oviii}{O~{\sc viii}}
\newcommand{\heii}{He~{\sc ii}}

\begin{document}

\thesaurus{3(11.01.2; 11.09.1; 11.17.1; 11.19.1; 13.21.1; 13.25.2)}

\title{Multiwavelength studies of the Seyfert 1 galaxy NGC7469. 
I - Far UV observations with FUSE}
\author{
G. A. Kriss\inst{1}\inst{2},
A. Blustin\inst{3},
G. Branduardi-Raymont\inst{3},
R. F. Green\inst{4},
J. Hutchings\inst{5},
M. E. Kaiser\inst{2}
}
\institute{
	Space Telescope Science Institute, 3700 San Martin Drive,
	Baltimore, MD, 21218, USA\\
	email: gak@stsci.edu
\and
	Center for Astrophysical Sciences, Johns Hopkins University,
	Baltimore, MD, 21218-2686, USA
\and
	Mullard Space Science Laboratory, University College London,
	Holmbury St. Mary, Dorking, Surrey, RH5 6NT, UK
\and
	Kitt Peak National Observatory,
        National Optical Astronomy Observatories, P.O. Box 26732,
        950 North Cherry Ave., Tucson, AZ, 85726-6732; rgreen@noao.edu
\and
	Dominion Astrophysical Observatory, National Research Council
	of Canada, Victoria, BC, V8X 4M6, Canada; john.hutchings@hia.nrc.ca
}

\offprints{G. A. Kriss}
\mail{G. A. Kriss, Space Telescope Science Institute, 3700 San Martin Drive,
Baltimore, MD 21218}

\date{Version 1.7 (Final version submitted to printers, 25 February 2003.)}

\titlerunning{FUSE spectrum of NGC 7469}
\authorrunning{Kriss et al.}

\maketitle

\begin{abstract}
We obtained far-ultraviolet spectra of the Seyfert 1 galaxy NGC~7469 using the
{\it Far Ultraviolet Spectroscopic Explorer} on 1999 December 6.
Our spectra cover the wavelength range 990--1187 \AA\ with a resolution of
$\sim 0.05$ \AA.
We see broad emission lines of \ciii, \niii, \ovi, and \heii\ as well
as intrinsic absorption lines in the \ovi\ $\lambda\lambda1032,1038$
resonance doublet.
The absorption arises in two distinct kinematic components at systemic
velocities of $-569$ \kms\ and $-1898$ \kms.
Both components are very highly ionized---no significant \lyb\ absorption
is present. The higher blueshift component is not quite saturated, and it
has a total \ovi\ column density of $8 \times 10^{14}~\rm cm^{-2}$.
It covers more than 90\% of the continuum and broad-line emission.
The lower blueshift component is heavily saturated and covers only $\sim50$\%
of the continuum and broad-line emission.
It too has a column density of $8 \times 10^{14}~\rm cm^{-2}$, but this is
less certain due to the high saturation.
We set an upper limit of $< 1.5 \times 10^{18}~\rm cm^{-2}$ on the \ovi\
column density of this component.  Its line depth is consistent with
coverage of only the continuum, and thus this component may lie interior
to the broad emission-line gas.
The component at $-569$ \kms\ has a velocity comparable to the high-ionization
X-ray absorption lines seen in the {\it XMM-Newton} grating spectrum of
NGC~7469, and photoionization models show that the observed column densities
of \ovi\ and \hi\ are compatible with their formation in the same gas
as that causing the X-ray absorption.
The gas at $-1898$ \kms\ has lower ionization and column density, and no
significant X-ray absorption is associated with it.
\end{abstract}


\section{Introduction\label{sec:intro}}

The Seyfert 1 galaxy NGC~7469 is one of a handful of bright, nearby active
galactic nuclei (AGN) that has been intensively studied at multiple
wavelengths
and monitored extensively in the X-ray, UV, and optical bands
(\cite{Wanders97};
\cite{Nandra98}; \cite{Collier98}; \cite{Kriss00a}).
ASCA X-ray spectra (\cite{Reynolds97}; \cite{George98}) were modeled with
absorption edges of {\sc O~vii} and {\sc O~viii} indicative of
a warm absorber, and {\it Hubble Space Telescope} Faint Object Spectrograph
observations show associated UV absorption lines in \lya, \nv\ and \civ.
Such X-ray and UV absorption is common in Seyfert 1s, occurring in tandem
in roughly half the population
(\cite{Reynolds97}; \cite{George98}; \cite{Crenshaw99}).
The X-ray and UV absorbing gas components thus appear to be related,
but it is not yet clear whether this is the consequence of a more general
phenomenon linking the gas visible at both wavelengths, or if the absorption
arises in the same gas with identical physical conditions in both the X-ray
and the UV.
At high resolution, both the X-ray and the UV absorption is kinematically
complex
(\cite{Kaastra01}; \cite{Kaspi00}; \cite{Kaspi02};
\cite{Crenshaw98}; \cite{Crenshaw99}; \cite{Mathur99}; \cite{Kriss00b}).
Generally only some of the UV absorption lines have velocities and
physical conditions similar to those seen in the X-ray.

The locations of both the X-ray and the UV absorbing media are also highly
uncertain.
If the outflowing gas originates in the accretion disk,
then it may be relatively close to the central engine
(e.g., \cite{Konigl94}; \cite{Murray95})
and interior to the broad emission-line region.
On the other hand, if the absorbing gas is a thermal wind driven off the
obscuring torus (\cite{KK95}; \cite{KK01}), then it will lie at distances
of roughly 1 pc, be external to the broad-line region, but probably interior
to the narrow-line region.
Studies of the X-ray absorption variability (NGC~3516; \cite{Netzer02})
show that at least some of the gas does lie at distances approaching 1 pc.
Variability of UV absorption in NGC~4151 suggests distances of tens of
parsecs for the absorbing gas (\cite{Espey98}), and even greater distances
for the associated UV absorption seen in some quasars (\cite{Hamann95};
\cite{Hamann97}).

Using the {\it Far Ultraviolet Spectroscopic Explorer (FUSE)}, we obtained
high-resolution UV spectra below 1200 \AA\ covering the
\ovi\ $\lambda\lambda1032,1038$ resonance doublet and the high-order Lyman
lines.
The \ovi\ doublet provides a key link to the X-ray band since high energy
transitions from the same ion can be viewed with {\it XMM-Newton}.
This permits us to directly compare the kinematics and the column densities
measured in both wavelength regions as a crucial test for establishing the
link between X-ray and UV absorbing gas in Seyferts.
{\it XMM-Newton} observations of NGC~7469 are discussed in a
companion paper (\cite{Blustin03}).

\section{Observations\label{sec:obs}}

\FUSE\ observes the far-ultraviolet wavelength range from 912--1187 \AA\
using
four independent optical channels.
In each channel a primary mirror gathers light for a Rowland-circle
spectrograph. Two-dimensional photon-counting detectors record the
dispersed spectra. Two of the optical trains use LiF coatings on the
optics to cover the 990--1187 \AA\ wavelength range.
The other two channels cover shorter wavelengths down to 912 \AA\ using
SiC-coated optics.  See \cite{Moos2000} for a
full description of \FUSE\ and its in-flight performance.

We observed NGC~7469 with \FUSE\ on 1999 December 6 through the
30\arcsec-square
low-resolution apertures.
Data were obtained on 22 consecutive orbits for a total integration time
of 37,803 s.
Unfortunately the SiC channels were not
properly aligned; the only detectable flux was in the LiF channels.
The time-tagged data were processed using v1.8.7 of the \FUSE\ calibration
pipeline.
\cite{Sahnow2000} describe the standard \FUSE\ data
processing steps. We added an additional customized step to normalize and
subtract a flat background image from each detector segment.
The resulting extracted spectra were merged into a linearized spectrum with
0.05 \AA\ bins. While this rebinning process introduces some correlated
error,
since each of these bins holds $\sim7$ original pixels, the resulting bins
are more than 80\% statistically independent.
Poisson errors and data quality flags are propagated through the data
reduction
process along with the science data.

To firmly establish the zero-point of our wavelength scale, we compare the
positions of low-ionization Galactic absorption lines from species such as
Ar~{\sc i}, Fe~{\sc ii}, {\sc O~i} and $\rm H_2$ to the Galactic
21 cm \hi\ velocity as measured by Murphy et al. (unpublished).
This requires a shift of $-0.32$ \AA\ to be applied to our wavelengths
to place them in a heliocentric reference frame.
We estimate that the flux scale is accurate to $\sim$10\%, and that wavelengths
are accurate to $\sim$15 \kms.

\begin{figure*}
\psfig{figure=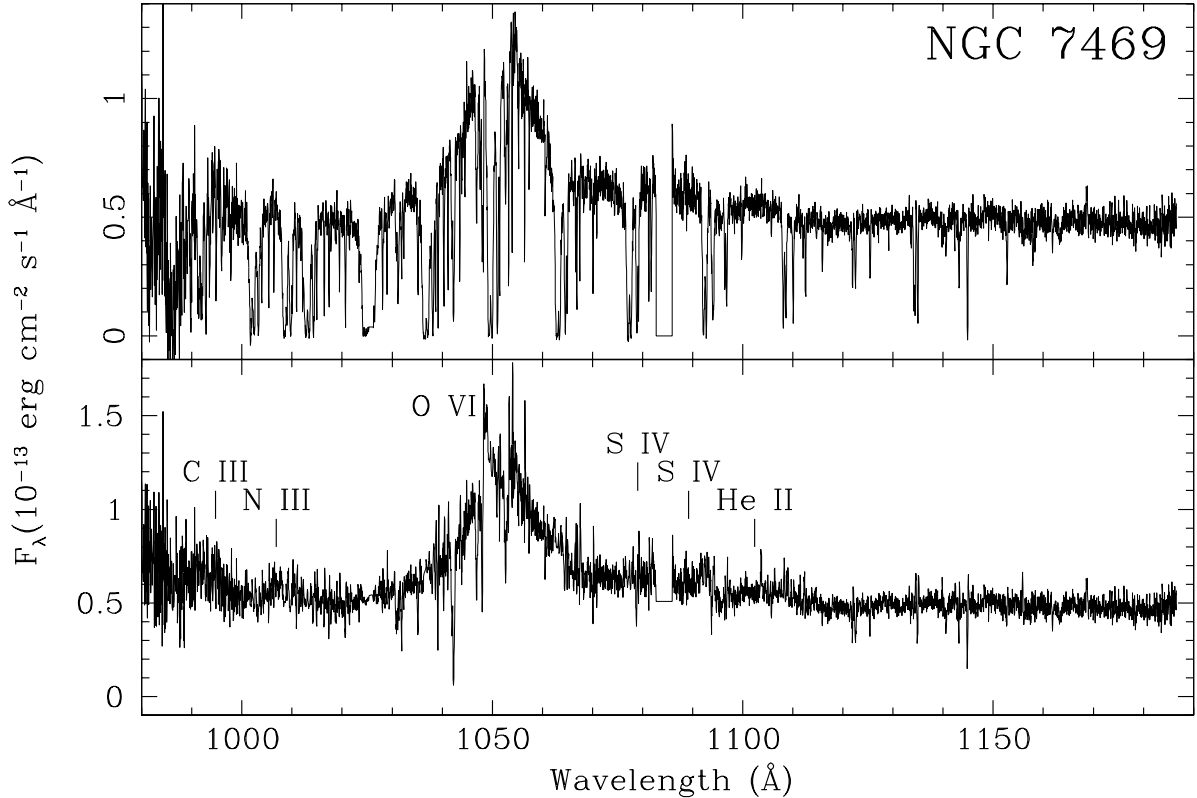,width=18.0cm,angle=-90}
\caption{
Upper panel: observed \FUSE\ spectrum of NGC 7469 with 0.05 \AA\ binning.
Lower panel: the \FUSE\ spectrum corrected for foreground interstellar
absorption.  Identified broad emission lines are marked.
\label{H4204F1.ps}}
\end{figure*}

The full merged spectrum is shown in the top panel of Fig. \ref{H4204F1.ps}.
Numerous Galactic absorption features obscure much of the spectrum, so we have
constructed a simple model of the interstellar absorption to remove them
from the spectrum.
We assume a Galactic \hi\ column of $4.4 \times 10^{20}~\rm cm^{-2}$ at a
heliocentric velocity of $-9$ \kms\ with a Doppler width of 10 \kms.
Heavy elements are included at solar abundances following the depletion
pattern typical of warm gas toward $\zeta$ Oph (\cite{SS96}).
Matching the molecular hydrogen absorption requires one component with
gas at 50 K and a $\rm H_2$ column density of $1 \times 10^{20}~\rm cm^{-2}$,
and a second component with a temperature of 300 K and a total column density
of $1 \times 10^{17}~\rm cm^{-2}$.
We show the spectrum of NGC~7469 corrected for this foreground absorption
in the lower panel of Fig. \ref{H4204F1.ps}.
Strong, broad \ovi\ emission dominates the spectrum. Weaker broad emission
lines of \ciii\ $\lambda$977, \niii\ $\lambda$991, and \heii\ $\lambda1085$
are also visible, as well as several unidentified emission features.
The low broad bump on the red wing of \ovi\ is also present in the \FUSE\
spectrum of Mrk~509 (\cite{Kriss00b}) as well as in the spectra of low-redshift
quasars observed using the {\it Hubble Space Telescope} (\cite{Laor95}).
Based on \FUSE\ observations of a large sample of low-redshift AGN,
Kriss et al. (2003) have identified this as blended emission due to
{\sc S~iv} $\lambda\lambda 1062,1072$.

\begin{figure}
\psfig{figure=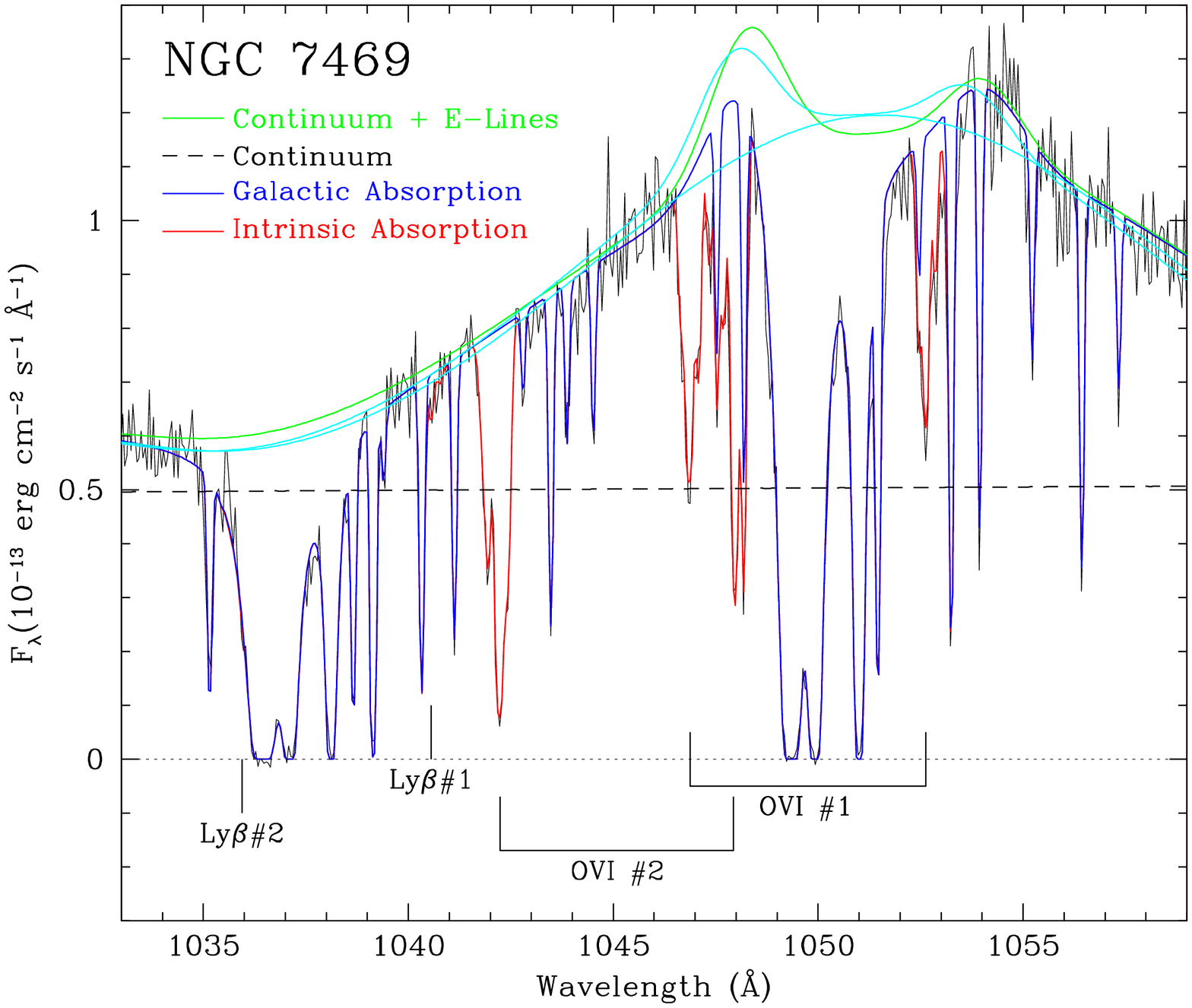,width=8.75cm,angle=-90}
\caption{
FUSE spectrum of NGC 7469 (with 0.05 \AA\ bins) in the Ly$\beta$/\ovi\
region
is shown as the thin black line.  The overlayed color lines show the best
fit (model A) as described in the text.
The thin green line represents the fitted continuum and emission
components. The thin red line shows the fitted intrinsic absorption
components.  The thin blue line shows the foreground Galactic absorption lines.
The light blue lines illustrate the continuum and narrow O VI line profiles
for the two extreme cases we considered, models D and E,
which have no narrow line emission, or 2:1 narrow-line emission that
is not absorbed, respectively.
The intrinsic absorbing gas is readily visible as two separate complexes
outlined in red.
Corresponding absorption in Ly$\beta$ is weak at best.
\label{H4204F2.ps}}
\end{figure}

At the high resolution of \FUSE, absorption features intrinsic to NGC~7469 are
also discernible. Fig. \ref{H4204F2.ps} shows the \lyb\ and \ovi\ region of
the spectrum in 0.05 \AA\ bins. Here one can see two absorption systems due
to the \ovi\ $\lambda\lambda$1032,1038 resonance doublet in the blue wing
of the broad \ovi\ emission line. The \lyb\ absorption associated with
these systems is weak, if present at all, and severely blended with
foreground Galactic absorption. Similarly, no \lyg\ or \ciii\ $\lambda$977
absorption is visible at shorter wavelengths.

\section{Analysis\label{sec:analysis}}

To measure the emission and absorption line properties of NGC~7469, we
model the spectrum using the IRAF task {\tt specfit} (\cite{Kriss94}).
We use a reddened power law in $f_{\lambda}$ to describe the continuum
emission.
For extinction, we use a \cite{CCM89}
curve with $R_V = 3.1$ and E(B$-$V)=0.069 (\cite{Schlegel98}).
To model the continuum, we use regions of the spectrum free of both emission
and absorption lines covering the full observed wavelength range.
The best-fit, extinction-corrected continuum has the form
$f_{\lambda} =\rm (1.34 \pm 0.02) \times 10^{-13} (\lambda / 1000
\AA)^{-1.25 \pm 0.10}~\ergscmA$.

To model the \ovi\ emission and absorption, we fix the continuum parameters
as determined above and restrict our subsequent fits to the
1033.0--1060.3 \AA\ wavelength range.
As one can see from the models shown in Fig. \ref{H4204F2.ps},
the \ovi\ emission line profile requires both broad and narrow components.
We model these with Gaussian line profiles.
The narrow component is only cleanly visible near the peak of the red
component of the \ovi\ doublet; strong foreground Galactic $\rm H_2$
absorption obscures the corresponding feature near the blue peak.
Although we can only clearly see part of the red portion of the
narrow doublet, our baseline model assumes that they are optically thin
with a 2:1 intensity ratio. (Tests using other assumptions are described in
our discussion of the absorption line fits below.)
The broad component is also treated as a doublet with two blended Gaussians
having a 2:1 intensity ratio.
For each narrow and broad doublet, we require the widths to be identical
and the wavelengths to have the same ratio as their laboratory values.
The \ciii\ $\lambda977$, \niii\ $\lambda991$,
{\sc S~iv} $\lambda\lambda 1062,1072$,
and \heii\ $\lambda1085$
emission lines are all modeled as single Gaussian emission lines.
The two components of the {\sc S~iv} doublet have identical widths, a fixed
flux ratio of 1:1, and wavelengths fixed at the ratio of their laboratory value.
In addition, we include a single Gaussian emission line at the location
of the unidentified emission feature at 1031 \AA.
Table \ref{tbl-elines} gives the best-fit parameters for these emission lines.

\begin{table*}
\footnotesize
\caption[]{Emission lines in NGC~7469}\label{tbl-elines}
\begin{tabular}{lcccccc}
\hline
\noalign{\smallskip}
Line & $\rm \lambda_{vac}$ & Flux$^{\rm a}$ & Velocity$^{\rm b}$ & FWHM \\
    & (\AA) & & $\rm ( km~s^{-1} )$ & $\rm ( km~s^{-1} )$ \\
\noalign{\smallskip}
\hline
\noalign{\smallskip}
C III      & 977.02 & $\phantom{0}9.5 \pm 2.1$ & $\phantom{-}271 \pm 160$ &
$1599 \pm 210$ \\
N III      & 991.58 & $\phantom{0}5.3 \pm 1.3$ & $\phantom{-0}95 \pm 158$ &
$1599 \pm 210$ \\
Unknown    & 1014.86 & $\phantom{0}6.5 \pm 1.1$ & $\phantom{-00}0 \pm 160$ &
$1939 \pm 518$ \\
O VI broad  & 1031.93 & $85.8 \pm 4.3$ & $\phantom{-}323 \pm \phantom{0}37$
& $4901 \pm 146$ \\
O VI broad  & 1037.62 & $42.9 \pm 2.2$ & $\phantom{-}323 \pm \phantom{0}37$
& $4901 \pm 146$ \\
O VI narrow & 1031.93 & $23.5 \pm 2.9$ & $-163 \pm \phantom{0}55$
& $1061 \pm \phantom{0}82$ \\
O VI narrow & 1037.62 & $11.8 \pm 1.5$ & $-163 \pm \phantom{0}55$
& $1061 \pm \phantom{0}82$ \\
S IV       & 1062.66 & $15.5 \pm 1.3$ & $\phantom{-}194 \pm 142$ & $7637 \pm 251$ \\
S IV       & 1072.97 & $15.5 \pm 1.3$ & $\phantom{-}194 \pm 142$ & $7637 \pm 251$ \\
HeII       & 1085.15 & $\phantom{0}3.2 \pm 0.3$ & $\phantom{0}$$-47 \pm 161$
& $2575 \pm 210$ \\
\noalign{\smallskip}
\hline
\end{tabular}
\begin{list}{}{}
\item[$^{\mathrm{a}}$]
Observed flux in units of $\rm 10^{-14}~erg~cm^{-2}~s^{-1}$.
\item[$^{\mathrm{b}}$]
Velocity is relative to a systemic redshift of
$cz = 4916~\rm km~s^{-1}$ \cite{RC3}.
\end{list}
\end{table*}

We model absorption lines in the NGC~7469 spectrum using Voigt profiles.
For the intrinsic absorbers, we permit the absorption to partially
cover the source.  That is, for a covering fraction $f_c$, the
transmittance at a given wavelength, $t(\lambda)$, has the form
$t(\lambda) = 1 + f_c (e^{-\tau(\lambda)} - 1)$, where $\tau(\lambda)$ is
the optical depth at that wavelength.
The intrinsic absorption lines are clustered into two complexes
that we designate as \#1 (at a relative systemic velocity of $-569~\kms$)
and \#2 (at velocity $-1898~\kms$).
Each of these complexes is modeled with four separate blended absorption
lines. The \ovi\ lines in each component are treated as doublets with
their wavelengths fixed at the ratio of their laboratory values, their
optical depths fixed at a 2:1 ratio, and their velocity widths forced to
be the same. For the corresponding \lyb\ lines, we fix their wavelengths
at the ratio of their laboratory values to those of the \ovi\ lines and
force them to have the same velocity widths. Covering fractions vary
freely within each complex. The individual \ovi\ lines
within a complex are forced to have the same velocity widths and the
same covering fractions. (This assumption is relaxed below in an
alternative model for the absorption.) Since \lyb\ absorption is
not even unambiguously detectable, we fix the relative optical depths of
the lines within each complex to be the same as the ratios observed for
\ovi. However, we permit the overall optical depth for each \lyb\ complex
to vary freely.
Table \ref{tbl-alines} gives the best-fit parameters for this model of the
absorption lines.
Parameters in the table with error bars of zero had their values linked
to another parameter in the fit.

\begin{table*}
\footnotesize
\caption{Absorption lines in NGC~7469\label{tbl-alines}}
\begin{tabular}{lccccccc}
\hline
\noalign{\smallskip}
Feature & Comp & $\lambda_{\mathrm{vac}}$ & $W^{\mathrm{a}}_\lambda$ &
$\mathrm N_{ion}$ & $\Delta \mathrm v^{\mathrm{b}}$ & FWHM & $f_c$ \\
     & \#    & (\AA)  & (\AA) & $\rm ( 10^{12}~cm^{-2} )$ &
($\rm km~s^{-1}$) & ($\rm km~s^{-1}$) &  \\
\noalign{\smallskip}
\hline
\noalign{\smallskip}
Ly$\beta$ & 2a & 1025.72 & $<$0.05 & $\phantom{00}1.5 \pm \phantom{00}3.4$ & $-2016
\pm 12$ & $29 \pm 1$ & $0.93 \pm 0.03$ \\
Ly$\beta$ & 2a & 1025.72 &  & $\phantom{00}3.2 \pm \phantom{00}7.3$ & $-1975
\pm \phantom{0}4$ & $29 \pm 1$ & $0.93 \pm 0.03$ \\
Ly$\beta$ & 2a & 1025.72 &  & $\phantom{0}12.9 \pm \phantom{0}29.5$ & $-1898
\pm \phantom{0}1$ & $29 \pm 1$ & $0.93 \pm 0.03$ \\
Ly$\beta$ & 2a & 1025.72 &  & $\phantom{00}4.0 \pm \phantom{00}9.1$ & $-1840
\pm \phantom{0}3$ & $29 \pm 1$ & $0.93 \pm 0.03$ \\
Ly$\beta$ & 1a & 1025.72 & $<$0.03 & $\phantom{00}8.2 \pm \phantom{00}4.3$ &
$\phantom{0}$$-627 \pm \phantom{0}5$ & $25 \pm 2$ & $0.53 \pm 0.05$ \\
Ly$\beta$ & 1b & 1025.72 &  & $\phantom{0}53.2 \pm \phantom{0}28.1$ &
$\phantom{0}$$-569 \pm \phantom{0}4$ & $25 \pm 2$ & $0.53 \pm 0.05$ \\
Ly$\beta$ & 1c & 1025.72 &  & $\phantom{0}15.3 \pm \phantom{00}8.1$ &
$\phantom{0}$$-506 \pm \phantom{0}3$ & $25 \pm 2$ & $0.53 \pm 0.05$ \\
Ly$\beta$ & 1d & 1025.72 &  & $\phantom{00}3.9 \pm \phantom{00}2.1$ &
$\phantom{0}$$-431 \pm \phantom{0}5$ & $25 \pm 2$ & $0.53 \pm 0.05$ \\
OVI     & 2a & 1031.93 & $0.48 \pm 0.02$ & $\phantom{0}36.1 \pm \phantom{0}16.2$ & $-2015
\pm 12$ & $29 \pm 1$ & $0.93 \pm 0.03$ \\
OVI     & 2b & 1031.93 &  & $121.2 \pm \phantom{0}16.3$ & $-1973 \pm
\phantom{0}4$ & $29 \pm 1$ & $0.93 \pm 0.03$ \\
OVI     & 2c & 1031.93 &  & $494.0 \pm \phantom{0}51.4$ & $-1898 \pm
\phantom{0}1$ & $29 \pm 1$ & $0.93 \pm 0.03$ \\
OVI     & 2d & 1031.93 &  & $154.4 \pm \phantom{0}16.9$ & $-1838 \pm
\phantom{0}3$ & $29 \pm 1$ & $0.93 \pm 0.03$ \\
OVI     & 1a & 1031.93 & $0.23 \pm 0.02$ & $\phantom{0}76.1 \pm \phantom{0}25.3$ &
$\phantom{0}$$-626 \pm \phantom{0}5$ & $25 \pm 2$ & $0.53 \pm 0.05$ \\
OVI     & 1b & 1031.93 &  & $494.7^{+1.5\times10^6}_{-167.2}$ &
$\phantom{0}$$-569 \pm \phantom{0}4$ & $25 \pm 2$ & $0.53 \pm 0.05$ \\
OVI     & 1c & 1031.93 &  & $142.3 \pm \phantom{0}28.2$ & $\phantom{0}$$-506
\pm \phantom{0}3$ & $25 \pm 2$ & $0.53 \pm 0.05$ \\
OVI     & 1d & 1031.93 &  & $\phantom{0}58.7 \pm \phantom{0}15.0$ &
$\phantom{0}$$-432 \pm \phantom{0}5$ & $25 \pm 2$ & $0.53 \pm 0.05$ \\
\noalign{\smallskip}
\hline
\end{tabular}
\begin{list}{}{}
\item[$^{\mathrm{a}}$]
Equivalent widths are integrated over all components of each spectral feature.
Upper limits are $2 \sigma$.
\item[$^{\mathrm{b}}$]
Velocity is relative to a systemic redshift of
$cz = 4916~\rm km~s^{-1}$ \cite{RC3}.
\end{list}
\end{table*}

Since much of the peak of the \ovi\ emission-line profile is obscured by
foreground Galactic absorption, we have explored the effects that this
uncertainty in the true emission-line profile might have on our
characterization of the absorption lines.
We consider a wide range of models:

\noindent
A.
This is our baseline model, described in detail above.
The narrow emission lines have line ratios fixed at an optically thin value
of 2:1, and we assume that this narrow line emission is obscured
at the same covering fraction as the continuum and broad emission line.
Individual \ovi\ lines within a complex have the same velocity widths
and covering fractions.

\noindent
B.
In this model the narrow emission lines are treated the same as in ``A",
but the velocity widths and the covering fractions of individual \ovi\
lines within a complex are all permitted to vary freely.
In these fits, the strongest, deepest absorption line in each complex
has parameters typical of those found in our baseline model.
The weaker adjacent lines retain similar widths in these fits, but
their best-fit covering fractions do change.  However, the individual
covering fractions are not well constrained. Their error bars are
typically $\pm 20$\%, and the new values lie within one error bar of
the best-fit value in the baseline model.

\noindent
C.
This variation has the narrow emission line fluxes fixed at an optically
thick ratio of 1:1.
The emission is obscured with the same covering fraction as the continuum
and broad emission line.
As in the baseline model, individual \ovi\ lines within a complex have
the same velocity widths and covering fractions. In a pattern that repeats
with each of the subsequent variations below, the best-fit \ovi\ column
densities in this model are slightly lower, but not more than twice the
values of the error bars shown in Table \ref{tbl-alines}.
While this model significantly changes the overall emission-line profile,
since the \ovi\ absorption is so highly blueshifted, little of the
intrinsic absorption falls in the region of the line profile with the
most dramatic changes.  Thus the overall optical depth of the absorption
lines is roughly the same as in the baseline model.

\noindent
D.
We also considered a model with no narrow emission-line component.
As in the baseline model, individual \ovi\ lines within a complex have
the same velocity widths and covering fractions.
This model provides a significantly worse fit than the baseline,
but the measured \ovi\ column densities and covering fractions
are the same as the baseline model to within the error bars in
Table \ref{tbl-alines}.

\noindent
E.
This model has narrow-line emission with line ratios fixed at 2:1,
but here we assume that the intrinsic
absorption does not obscure the narrow-line emission at all.
As in the baseline model, individual \ovi\ lines within a complex have
the same velocity widths and covering fractions.
Again the \ovi\ column densities are lower, but they are not significantly
different from those of the baseline model.

\noindent
F.
In this model we also assume that the intrinsic absorption does not absorb
the narrow-line emission, but we fix that emission at a ratio of 1:1.
As in the baseline model, individual \ovi\ lines within a complex have
the same velocity widths and covering fractions.
Again we find no significant variation from the results of our baseline
model.

The overall properties of these various emission-line profile models are
summarized in Table \ref{tbl-models1}.
For the \ovi\ components 1 and 2 we list the total column density from the
fit, the velocity of the strongest feature in each component, the best-fit
widths of the absorption lines, and the best-fit covering fractions.
The comparative summary for component \#1 is listed first in the table,
followed by the summary for component \#2.

\begin{table*}
\caption{Summary of properties of absorber models for
NGC~7469\label{tbl-models1}}
\begin{tabular}{l c c c c c}
\hline
\noalign{\smallskip}
Model/Component & $\chi^2 / \nu$ & $\rm N_{OVI}$ &
$\Delta \rm v^{\mathrm{a}}$ & FWHM & $f_c$ \\
 & & $\rm (10^{12}~cm^{-2} )$ &
($\rm km~s^{-1}$) & ($\rm km~s^{-1}$) &  \\
\noalign{\smallskip}
\hline
\noalign{\smallskip}
A/1 &  604.0/344 & $\phantom{0}772 \pm 172$ & $\phantom{0}$$-569 \pm 4$ &
$25 \pm 2$ & $0.53 \pm 0.05$\\
B/1 &  596.9/332 & $\phantom{0}797 \pm 317$ & $\phantom{0}$$-572 \pm 5$ &
$24 \pm 6$ & $0.54 \pm 0.06$\\
C/1 &  613.1/344 & $\phantom{0}820 \pm 217$ & $\phantom{0}$$-569 \pm 4$ &
$24 \pm 2$ & $0.51 \pm 0.04$\\
D/1 &  747.5/344 & $\phantom{0}933 \pm 244$ & $\phantom{0}$$-572 \pm 3$ &
$24 \pm 2$ & $0.50 \pm 0.04$\\
E/1 &  622.1/344 & $\phantom{0}758 \pm 163$ & $\phantom{0}$$-569 \pm 3$ &
$25 \pm 2$ & $0.56 \pm 0.05$\\
F/1 &  616.1/344 & $\phantom{0}815 \pm 195$ & $\phantom{0}$$-569 \pm 4$ &
$24 \pm 2$ & $0.54 \pm 0.04$\\
A/2 &  604.0/344 & $\phantom{0}806 \pm \phantom{0}59$ & $-1898 \pm 1$ & $29
\pm 1$ & $0.93 \pm 0.03$\\
B/2 &  596.9/332 & $\phantom{0}795 \pm 233$ & $-1898 \pm 3$ & $30 \pm 5$ &
$0.93 \pm 0.03$\\
C/2 &  613.1/344 & $\phantom{0}750 \pm \phantom{0}55$ & $-1898 \pm 1$ & $29
\pm 1$ & $0.95 \pm 0.03$\\
D/2 &  747.5/344 & $\phantom{0}680 \pm \phantom{0}58$ & $-1898 \pm 1$ & $29
\pm 1$ & $0.97 \pm 0.04$\\
E/2 &  622.1/344 & $1020 \pm \phantom{0}96$ & $-1895 \pm 1$ & $27 \pm 1$ &
$0.90 \pm 0.02$\\
F/2 &  616.1/344 & $\phantom{0}993 \pm \phantom{0}98$ & $-1895 \pm 1$ & $27
\pm 1$ & $0.90 \pm 0.02$\\
\noalign{\smallskip}
\hline
\end{tabular}
\begin{list}{}{}
\item[$^{\mathrm{a}}$]
Velocity is relative to a systemic redshift of
$cz = 4916~\rm km~s^{-1}$ \cite{RC3}.
\end{list}
\end{table*}

\begin{figure}
\psfig{figure=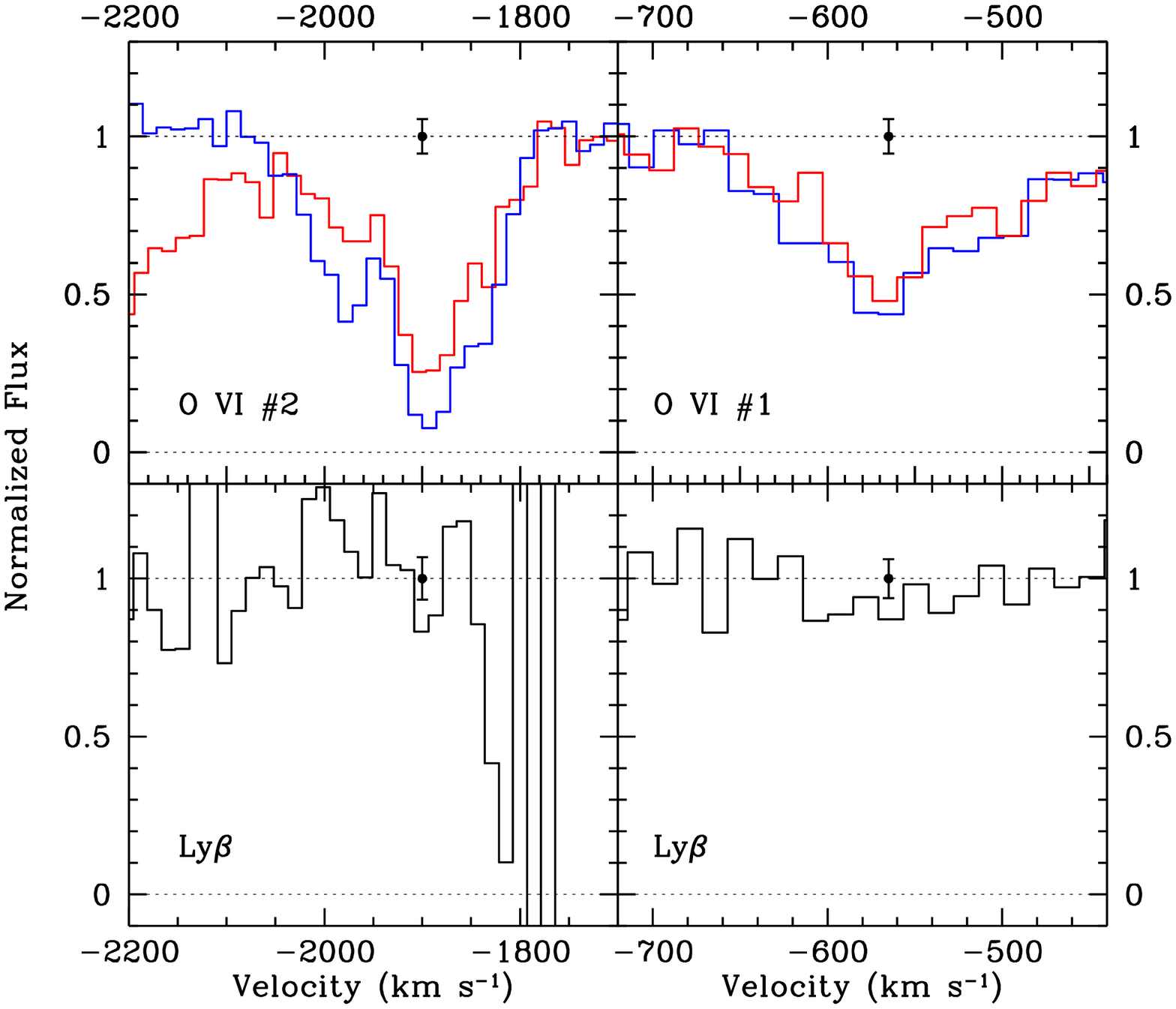,width=8.75cm,angle=-90}
\caption{
The normalized line profiles in the upper panels show intrinsic
{\sc O VI} $\lambda 1032$ (the solid blue line) and
{\sc O VI} $\lambda 1038$ (the solid red line) absorbing components 1 and 2.
These data are binned at 0.05 \AA, and foreground Galactic
absorption features have been divided out.
The normalized profiles in the lower panels show the corresponding location
in velocity space of expected Ly$\beta$ absorption, which is not
definitively
detected.
Representative error bars at the velocity centroid of each component are
shown.
\label{H4204F3.ps}}
\end{figure}

Fig. \ref{H4204F2.ps} graphically illustrates our best fit to the \ovi\
and \lyb\ region of the \FUSE\ spectrum.
Using our model for the underlying emission and the
foreground Galactic absorption, we divide this into the data to produce a
normalized spectrum that is corrected for the effects of foreground
Galactic absorption.
This normalized spectrum is shown in velocity space surrounding the
two absorption line complexes in \ovi\ and \lyb\ in
Fig. \ref{H4204F3.ps}.
Since the relative oscillator strengths of the blue and red lines in the
\ovi\ doublet have roughly a 2:1 ratio, one would expect the red line to
have a depth that is half that of the blue line in optically thin gas,
and an equal depth in optically thick gas.
From the normalized spectrum it is readily apparent that the equal depths
of the red and blue lines in Component \#1 imply that the absorption
is heavily saturated.
However, since the line cores are not black at their centers, we also can
see that the absorbing gas does not fully cover the underlying emission.
The covering fraction is only $\sim50$\%.
Our fits may underestimate the total column density in this absorption
complex. The lack of visible damping wings on the line profile
permits us to set an upper limit of $1.5 \times 10^{18}~\rm cm^{-2}$
on the \ovi\ column of this component. This upper limit is reflected in
the error bar that we assign to the strongest line in Component \#1 in
Table \ref{tbl-alines}.
The relative depths of the red and blue lines in Component \#2 are not
quite equal, but they also are not at an optically thin ratio.
This component is approaching saturation, but is not yet optically thick,
and so its total column density is better determined.

\section{Discussion\label{sec:discuss}}

As typical for Seyfert 1 galaxies viewed at high spectral resolution in the
far-ultraviolet, we see complex, multiple absorption lines from highly
ionized \ovi.
Although \lyb\ absorption in the \FUSE\ spectrum is weak or absent, there are
\lya\ absorption lines in the FOS spectrum of NGC~7469 obtained in 1996 at
systemic velocities of $-1870 \pm 17~\kms$ and $-656 \pm 24~\kms$
(\cite{Kriss00a}), very close to the velocities of
Components \#1 and \#2, so there is some minimal neutral hydrogen column.
The high \ovi\ to \hi\ column-density ratios for both components
imply very highly ionized gas consistent with those typical of
X-ray warm absorbers and similar to the highest ionization component
in Mrk~509 (\cite{Kriss00b}).
However, only Component \#1 has
velocities comparable to those of higher ionization states of \ovii\ and
\oviii\ ($-900 \pm 100~\kms$) as seen in the {\it XMM-Newton} grating spectrum
of NGC~7469 (\cite{Blustin03}). The upper limit on the \ovi\ column density
measured in the {\it XMM-Newton} grating spectrum of $< 10^{16.5}~\rm cm^{-2}$
is consistent with the column densities measured in the \FUSE\ spectrum.

\begin{table*}
\caption{Photoionization models for the absorbers in NGC~7469\label{tbl-models2}}
\begin{tabular}{l c c c c c c c}
\hline
\noalign{\smallskip}
$U$ & $\rm log~N_{tot}$ & $\rm N(HI)$ & $\rm N(OVI)$ & $\rm N(OVII)$ & $\rm N(OVIII)$ & $\rm N(CIV)$ & $\rm N(NV)$ \\
 & $\rm ( cm^{-2} )$ & $\rm ( cm^{-2} )$ & $\rm ( cm^{-2} )$ & $\rm ( cm^{-2} )$ \\
\noalign{\smallskip}
\hline
\noalign{\smallskip}
6.0$^{\rm a}$  & 20.55 & $1.1\times10^{14}$ & $8.0\times10^{14}$ & $7.3\times10^{16}$ & $1.3\times10^{17}$ & $1.1\times10^{12}$ & $1.0\times10^{13}$ \\
              & Observed: & $(8.0 \pm 4.2)\times10^{14}$ & $(8.0 \pm 3.0) \times10^{14}$ & $(3^{+5}_{-3})\times10^{16}$ & $(2^{+2}_{-1})\times10^{17}$ & \ldots & \ldots \\
0.2$^{\rm b}$ & 18.58 & $6.9\times10^{13}$ & $5.5\times10^{14}$ & $8.2\times10^{14}$ & $2.0\times10^{13}$ & $2.4\times10^{13}$ & $4.3\times10^{13}$ \\
              & Observed: & $(2.1 \pm 4.8)\times10^{13}$ & $(8.0 \pm 2.3) \times10^{14}$ & $<3\times10^{16}$ & $<3 \times 10^{16}$ & \ldots & \ldots \\
0.08$^{\rm c}$& 18.58 & $2.0\times10^{14}$ & $6.7\times10^{14}$ & $8.7\times10^{14}$ & $2.9\times10^{12}$ & $5.7\times10^{13~\rm d}$ & $7.2\times10^{14~\rm e}$ \\
\noalign{\smallskip}
\hline
\end{tabular}
\begin{list}{}{}
\item[$^{\mathrm{a}}$]
This model is our best match to Component \#1.
\item[$^{\mathrm{b}}$]
This model is our best match to Component \#2.
\item[$^{\mathrm{c}}$]
This model corresponds to Component \#2 at the epoch of the 1996 FOS observations.
\item[$^{\mathrm{d}}$]
For a Doppler width of 100 \kms, the blue line in the doublet would have
an equivalent width of 0.20 \AA.
\item[$^{\mathrm{e}}$]
For a Doppler width of 100 \kms, the blue line in the doublet would have
an equivalent width of 0.16 \AA.
\end{list}
\end{table*}

\begin{figure}
\psfig{figure=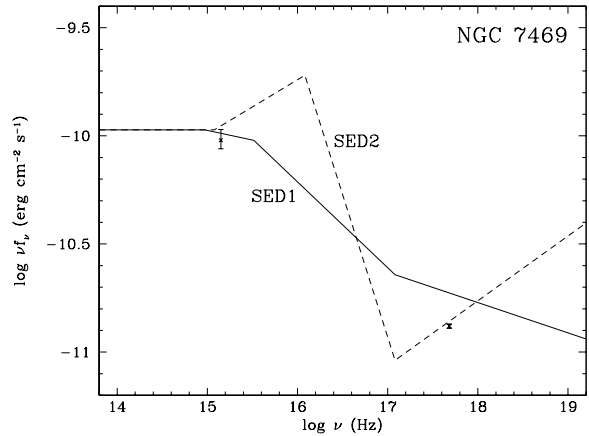,width=8.75cm,angle=-90}
\caption{
Spectral energy distributions used for photoionization models of absorbing gas
in NGC~7469. SED1, based on Kriss et al. (2000), is the solid line.
SED2, based on the spectrum in this paper, is the dashed line.
The curves are normalized so that SED2 matches the extinction-corrected flux at 1000 \AA\ in the FUSE spectrum.
The points with error bars showing the UV and X-ray fluxes as observed with
{\it XMM-Newton} are consistent with SED2.
\label{H4204F4.ps}}
\end{figure}

To test whether the same gas is responsible for both
the UV and the X-ray absorption in component \#1, we computed photoionization
models similar to those used by Krolik \& Kriss (1995, 2001).
These models cover a grid
in total column density ranging from $\rm N_{tot} = 10^{18}$ to
$10^{21}~\rm cm^{-2}$, and in ionization parameter from $U = 0.05$ to 10.0,
where $U$ is the ratio of ionizing photons in the Lyman continuum to the
local electron density.
For the illuminating spectrum in our models, we used two different spectral
energy distributions.
The first, SED1, is the same as that described by Kriss et al. (2000a),
which is based on the simultaneous IUE, FOS, and RXTE observations of
NGC~7469 performed in 1996. We show SED1 as a solid line in
Fig. \ref{H4204F4.ps}.
The second spectral energy distribution, SED2, is based on the current \FUSE\ 
spectrum presented in this paper and the {\it XMM-Newton} spectrum from
Blustin et al. (2003).
At long wavelengths we assume $f_\nu \propto \nu^{-1}$, as in SED1.
At 2500 \AA, this breaks to $f_\nu \propto \nu^{-0.75}$ to match the \FUSE\ 
spectrum.
At X-ray energies, we use the energy index of 0.7 from the
{\it XMM-Newton} spectrum  to span the 0.5 keV to 100 keV.
At higher energies we use a steeper index  of 2.0 to prevent divergence
in the total ionizing flux.
From 50 eV to 0.5 keV, the steep $f_\nu \propto \nu^{-2.32}$ is a reasonable
match to the soft excess seen in the {\it XMM-Newton} data.
The \FUSE\ and {\it XMM-Newton} data were not obtained simultaneously, and so we normalize the X-ray
continuum relative to the UV continuum by choosing $\alpha_{ox} = 1.34$,
the same value used by Kriss et al. (2000a) based on the 1996 simultaneous
observations.
SED2 is the dashed line shown in Fig. \ref{H4204F4.ps}.
For comparison, we show the absorption-corrected UV and X-ray fluxes
at 2120 \AA\ and 2 keV, respectively, as
observed with {\it XMM-Newton} (\cite{Blustin03}).
These data have $\alpha_{ox} = 1.33 \pm 0.02$,
and they are consistent with SED2.

After examining the results of our modeling, we find that none of the models
using SED1 achieves sufficiently high ratios of \ovi, \ovii, and
\oviii\ relative to \hi\  to match the values observed in our spectra.
For SED2, we do find a solution that is compatible with our observations.
This model has U = 6.0 and log $N_{tot}$ = 20.55.  The predicted column
densities for ions of interest are summarized in Table \ref{tbl-models2}, where
we compare them to the observed column densities.
Note that \civ\ and \nv\ are virtually invisible in this model, and,
indeed, no absorption from these ions was seen by Kriss et al. (2000a) in the 
1996 FOS spectrum at the velocity of component \#1.

For Component \#2, the required column density and ionization level is
substantially lower.  We obtain a reasonable match to the observed \ovi\ and
\hi\  column densities for $U = 0.2$ and $N_{tot}$ = 18.58, as
also shown in Table \ref{tbl-models2}.
The predicted column densities for \civ\ and \nv\ in this model
are also rather low, although easily detectable.
We note that the UV continuum in the current \FUSE\ observation is a factor
of $2.1 \times$ brighter than that observed with the FOS in 1996.
If we assume that any variations since then are simply due to changes in flux
with a constant column density of absorbing gas, then we would have expected
the gas in 1996 to be responding to an ionization parameter of $U = 0.08$.
This would then predict \civ\ and \nv\ column densities of
$5.7 \times 10^{13}~\rm cm^{-2}$ and $7.2 \times 10^{13}~\rm cm^{-2}$,
respectively, with corresponding equivalent widths of the blue components
of each doublet (for optically thin gas) of 0.2 \AA\ and 0.16 \AA.
While these are still lower than the observed values in the FOS spectrum,
($0.45 \pm 0.05$ and $0.48 \pm 0.08$ \AA\ for \civ\ and \nv, respectively)
given the uncertainties in the spectral energy distribution and other
potentially variable factors, we find the agreement quite good.

The covering fraction and line depth of Component \#1 is consistent with its
absorbing only the continuum flux and none of the emission-line flux in
NGC~7469.  Its high ionization parameter, and its likely identification
with the same gas responsible for the X-ray absorption
suggests that it is much higher ionization than Component \#2.
Taken together, these points imply that Component \#1 lies very
close to the central engine, possible interior to the broad
emission line region. This would favor its association with an accretion disk
wind, rather than a thermally driven wind emanating from the obscuring torus.
In constrast, the high covering fraction of Component \#2 clearly places it
exterior to both the BLR and continuum regions as would be expected for the
thermally driven winds of Krolik \& Kriss (1995; 2001).
Overall, in this one object we may be seeing absorption from both
a torus wind and a disk wind.

\section{SUMMARY\label{sec:summary}}

Our high-resolution far-ultraviolet spectrum of NGC~7469 obtained with
\FUSE\ shows broad emission lines of \ciii, \niii, \ovi, and \heii, as well
as possible emission from {\sc S~iv} $\lambda\lambda 1062,1072$.
Intrinsic absorption in the \ovi\ $\lambda\lambda1032,1038$
resonance doublet arises in two distinct kinematic components at systemic
velocities of $-569$ \kms\ (Component \#1) and $-1898$ \kms\ (Component \#2).
Both components are very highly ionized with no significant \lyb\ absorption
detected at either velocity.
Component \#2, although highly ionized, has a lower total column density
than Component \#1, and it is consistent with having no associated X-ray
absorption.
It covers more than 90\% of both the continuum and broad-line emission.
Component \#1 at $-569$ \kms\ is the best match in velocity to the
highly ionized X-ray absorbing gas detected in the {\it XMM-Newton} grating
spectrum of NGC~7469 (\cite{Blustin03}) at a blueshift of $900 \pm 100~\kms$.
Photoionization models for Component \#1 show that for a total column
density of $10^{20.55}~\rm cm^{-2}$ and an ionization parameter of $U = 6.0$,
the column densities of \hi, \ovi, \ovii, and \oviii\ in the \FUSE\ and the
{\it XMM-Newton} spectra can all be reproduced.
Component \#1 also has an extraordinarily low covering fraction of ~0.5, and is
consistent with covering only the continuum emission and none of the broad-line
emission.  This suggests that it might arise in an accretion disk wind interior
to the broad line region.

\begin{acknowledgements}
This work is based on data obtained for the Guaranteed Time Team by the
NASA-CNES-CSA \FUSE\ mission operated by the Johns Hopkins University.
Financial support to U. S. participants has been provided by
NASA contract NAS5-32985.
The U. K. authors acknowledge the support of the Particle Physics and
Astronomy Research Council.
G. Kriss acknowledges additional support from NASA Long Term
Space Astrophysics grant NAGW-4443.
\end{acknowledgements}

\end{document}